\documentclass[aps,prl,twocolumn,showpacs]{revtex4}

\usepackage{graphicx}

\begin{document}

\title{Magnetic transitions in a Double Exchange-Holstein model with  e-ph
 interactions coupled to magnetism}

\author{L. G.  \surname{Sarasua} }
\email{sarasua@fisica.edu.uy}
\author{A.   \surname{Moreno-Gobbi} }

\affiliation{Instituto de F\'{i}sica, Facultad de Ciencias,  
Igu\'a 4225, CC 11400, Montevideo, Uruguay}

\author{M. A.   \surname{Continentino} }

\affiliation{Instituto de F\'{i}sica, Universidade Federal Fluminense, 
Campus da Praia Vermelha, Niter\'oi, RJ, Brasil}


\date{\today}

\begin{abstract}
In this work we study the Double Exchange-Holstein (DE-H) model
with an electron-phonon interaction $\gamma$  coupled to
magnetism. The analysis is performed combining  a mean-field
approximation for the double exchange interaction and the
Lang-Firsov transformation for the electron-phonon interaction.
Discontinuous magnetic transitions appear when the dependence
of $g$ with $m$ is sufficiently large, resembling  those
experimentally observed in manganites. We observe that the characteristic
resistivity peak that arises near the critical temperature appears
for  broad ranges of the system parameter values, unlike what occurs in a 
constant--$\gamma$ model.

\end{abstract}


\maketitle

\section{Introduction}

Perovskite manganites $La_{1-x}A_xMnO_3$ ($A=Ca, Sr, Ba$) have been studied 
intensively since the discovery of the spectacularly large dependence  of their 
resistivity with the applied magnetic field, a phenomenon called  colossal 
magnetoresistance (CMR).
This unusual high response to the application of a magnetic field occurs mainly  
for temperatures near the critical temperature  $T_c$ of  a ferromagnetic-metallic (FM) 
to  paramagnetic-insulating transition (PI) \cite{salamon}. 
In this range of temperatures, the curve of resistivity versus temperature exhibits 
a characteristic pronounced peak.
It was found that the ferromagnetic-to-paramagnetic transition might be 
of first or second order \cite{mira}, depending on the composition of the material.

The connection between ferromagnetism and metallic behavior can be explained with 
the Double Exchange (DE) model proposed by Zener \cite{zener}. This model assumes a 
very large Hund coupling between electrons and localized spins, which causes a 
reduction of the effective electron hopping when the ion spins are disordered \cite{anderson}.
Thus, the model predicts that the  system is  metallic in the ferromagnetic state and
an insulator in the paramagnetic one. 
However, the magnetoresistance effect that is obtained with this theory is very poor 
in comparison with the experimental observations, suggesting that the double exchange 
mechanism alone is not enough to explain the properties of the manganites \cite{millis}.
Millis, Littlewood and Shraiman \cite{millis} proposed  that a minimal model for 
the manganites must include a term describing a coupling between electrons and 
lattice degrees of freedom.
The effect of such interaction was considered by Roder, Zang and  Bishop  \cite{roder} , 
and  by Millis, Shraiman and Mueller \cite{millis2}. In Ref. \cite{roder} the authors 
showed that  the e-ph interaction gives a dependence of  $T_c$ with doping that is 
similar to the one observed experimentally.
In Ref. \cite{millis2}, it was found that the e-ph coupling enhances the  
magnetoresistance effect and does provide a resistivity behavior that is in 
qualitative agreement with the experimental observations. 
However, the resistivity peak near $T_c$ is obtained with this theory only if the 
electron-phonon coupling $\gamma$ satisfies the condition  $1.08 < \gamma  < 1.2 $. 
A similar result was obtained by Verg\'{e}s {\it et al} using Monte Carlo simulations 
\cite{verges}. The obtained phase diagram indicates that for the case $n=0.08$,  
there is a FM to PI transition if 1.3 $< \gamma < $ 1.6.

In a recent paper, Dagotto {\it et al} studied the resistivity of the DE model 
including e-ph interactions using Monte Carlo computational techniques \cite{dagotto}. 
In agreement with the previously mentioned works, it was found that the maximum  of
 resistivity is observed in the DE-H model if the coupling $\gamma$ is fine tuned 
around some values that depend on electron density $n$. For instance, for $n=0.1$, the 
maximum resistivity  appears for values of $\gamma$ in the range (1.3, 1.6), 
in agreement with the results of \cite{verges}.  The inclusion of disorder in the DE-H 
model was considered by Kumar and Majumdar in \cite{kumar} and also in \cite{dagotto}. 
It was found that quenched disorder favors polaronic formation and the appearance of 
the resistivity  peaks in the curve $\rho(T)$. However, it is not clear whether the
 magnetoresistance effect induced by disorder reproduces the experimentally observed 
resistive behavior in all their aspects.

 In this work we propose an alternative modification of the model. 
We study an DE-H with an electron-phonon coupling $\gamma$ that depends on  
the magnetic ordering, instead of being constant as it is considered in the usual model. 
The variation of $\gamma$ with disorder has been a long-standing problem. 
Both theoretical and experimental results show that $\gamma$ may have important 
changes induced by disorder \cite{sergeev, jan, zhong}. It is therefore of great 
interest to 
investigate how the magnetic transitions are affected by the variations of the 
e-ph coupling.  The work is organized as follows.  In section 2, after performing 
a Lang-Firsov transformation of the Hamiltonian,  we construct a free energy for 
the system following Kubo-Ohata theory. In section 3  we discuss   the dependence 
of the electron--phonon coupling with the magnetization. In section 4 we shall 
consider the effect of this dependence on the magnetic transitions and the resistivity. 
 In section 5 we summarize our results.

\section{Hamiltonian model}

The Hamiltonian of the Double Exchange-Holstein is given by

\begin{eqnarray}
& H & =  H_{Z} +H_{H} \nonumber \\ \nonumber \\
& H_{Z}&  = - \sum_{ij}t_{ij}(  c_{ i \sigma}^\dag c_{ j \sigma} +h.c. ) -J_H \sum_{i}   \sigma_i \cdot {\bf S}_i \nonumber \\
& H_H & = - \gamma \sum_{i} c_{ i \sigma}^\dag c_{ i\sigma} (b_{i}^\dag  + b_{ i}) + \omega  \sum_{i} ( b_{i}^\dag b_{i} +  \frac{1}{2} )
\label{H2}
\end{eqnarray}


\noindent where $t_{ij}$ is the electron hopping between the sites $i$ and 
$j$, $c_{ i \sigma}^\dag$ is the creation operator for the itinerant electrons, 
 $J_H$ is the Hund coupling, $\sigma_i$ is the spin of the itinerant electrons,
 $S_i$ is the spin of the $Mn$ ions, $\gamma$ is the electron-phonon coupling and
 $b_i^\dag$ are the phonon creation operators. Here we consider $S=3/2$, which is
 the total spin of the three $t_{2g}$ electrons of the $Mn^{+4}$ ions.  To study 
this Hamiltonian, we assume the usually considered regime  $J_H/t \rightarrow \infty $.  
It has been shown by various authors (\cite{kubo}, \cite{anderson}, \cite{mitra}), that
 in this limit $H_Z$  reduces to the DE Hamiltonian:

\begin{eqnarray}
H_{DE} = - \sum_{ij} t_{ij} \sigma_{ij} ( c_{ i \sigma}^\dag c_{ j \sigma} +h.c. )
\label{H2}
\end{eqnarray}
\noindent where
\begin{equation}
\sigma_{ij} = \langle \frac{S_T^{ij}+\frac{1}{2}}{2S+1} \rangle ,
\label{S2}
\end{equation}
$S_T^{ij}$ being the total spin of the subsystem formed by the ions at
 sites $i$,$j$ and the electron, i. e. 
$S_0^{ij}=| {\bf S}_i+{\bf S}_j+ {\bf \sigma}_i |$ \cite{kubo}, \cite{anderson}, 
\cite{mitra}. The value of $\sigma_{ij}$ must be obtained averaging over all the 
states of $S_T^{ij}$. To calculate different averages, we follow the mean-field 
approach of Kubo and Ohata  and introduce an effective field $\lambda = h_{eff} / T$ 
that tends to order the ion spins \cite{kubo},\cite{weibe}. The magnetization is 
obtained from

\begin{eqnarray}
&m(\lambda)& = z^{-1} \sum^S_{l=-S} (l/S) \exp(\lambda l/S),
\label{m}
\end{eqnarray}

\begin{eqnarray}
z(\lambda) =  \sum^S_{l=-S}  \exp(\lambda l/S).
\label{z}
\end{eqnarray}

 Since in the present model  $\gamma$ depends on the magnetization, it must be also a 
function of the field parameter $\lambda$. The value of   $\sigma_{ij}$ is also obtained 
averaging over the states of the dimer \cite{weibe}.  Substituting
 $\sigma_{ij}$ by it averaged value $\sigma$, the Hamiltonian (\ref{H2}) becomes

\begin{eqnarray}
 H  =  & -  \sigma (\lambda) &  \sum_{ij}t_{ij}( c_{ i \sigma}^\dag c_{ j \sigma} +h.c. )  - \gamma (\lambda) \sum_{i} c_{ i \sigma}^\dag c_{ i\sigma} (b_{i}^\dag  + b_{ i}) \nonumber \\ &+&  \omega \sum_{i} ( b_{i}^\dag b_{i} +  \frac{1}{2} ).
\label{H3}
\end{eqnarray}

To treat the phonon dependent part of (\ref{H3}) we use the Lang-Firsov transformation
 $\tilde{H} = e^{S} H e^{-S}$, where $S= g \sum{n_i (a_i-a_i^\dag)}$ and
 $g=\gamma / \omega$.
With this transformation, the Hamiltonian takes the form

\begin{eqnarray}
\tilde{H}  = - \sigma (\lambda) e^{-g^2} \sum_{ij} t_{ij} ( c_{ i \sigma}^\dag c_{ j \sigma} e^{-g(b_j^\dag-b_i^\dag)} e^{g(b_j-b_i)} +h.c. ) \nonumber  \\ - \omega g^2  \sum_{i} n_i  - 2 \omega g^2  \sum_{i} n_{ i \uparrow} n_{ i \downarrow} + \omega  \sum_{i} ( b_{i}^\dag b_{i} +  \frac{1}{2} ),
\label{H4}
\end{eqnarray}

\noindent where $n_{i \sigma}$ are the number operators  for the electrons and 
$n_{i} = n_{i \uparrow}+n_{i \downarrow}$. We approximate the wave function of 
the system as a tensorial product of  waves function for the electrons and phonons
 $|\psi \rangle = |\psi \rangle_e \otimes  |\psi_0 \rangle_{ph} $. Considering that
 $|\psi_0 \rangle_{ph}$ is the vacuum state and  averaging over this state, the
 Hamiltonian takes the form

\begin{eqnarray}
\tilde{H}  = & - & \sigma (\lambda) e^{-g^2(\lambda)}  \sum_{ij}t_{ij}(  c_{ i \sigma}^\dag c_{ j \sigma} +h.c.)  - \omega g^2(\lambda)  \sum_{i} n_i   \nonumber \\ &-& 2 \omega g^2(\lambda)  \sum_{i} n_{ i \uparrow} n_{ i \downarrow}.
\label{Hpro}
\end{eqnarray}

\noindent We note that the Lang-Firsov transformation introduces a Hubbard like
 attractive interaction which can promote the formation of bipolarons
 \cite{moskalenko, bonca}. However, we shall consider the system is in the 
 adiabatic limit 
$\omega \rightarrow 0 $ and thus we shall omit the last two terms of the above 
equation. 
From (\ref{Hpro}) we can obtain a free energy $F=E-TS$ for the system. 
The free energy of the electrons are given by

\begin{equation}
\Omega_e(\lambda) = -\frac{1}{2 \beta} \sum_{k , \sigma} \ln [1+e^{\beta (\mu - \epsilon_k(\lambda)) }],
\end{equation}

\noindent where $\epsilon_k = \sigma(\lambda) e^{-g^2(\lambda) } \epsilon_{0k}$,
and $\epsilon_{0k}$ stands by the energy levels of the bare band.
To obtain the free energy ${\cal F}$ of the whole system, we follow Kubo and Ohata
 \cite{kubo},  adding to $\Omega_e$ an entropy term corresponding to the spin ions.
 The free energy is thus given by
\begin{equation}
{\cal F}=\Omega_e(\lambda) - N T [ \ln z(\lambda) - \lambda m(\lambda)].
\label{F}
\end{equation}


In order to determine the mean-field solutions at given
temperature and doping level, we minimize ${\cal F}$ with respect to
the variational parameter $\lambda$. We still need, however, a
dependence of $g$ with $\lambda$, or equivalently, a dependence of
$g$ with $m$.

\section{Coupling between $g$ and the magnetization}

To justify the use of an electron-phonon parameter that varies with magnetization, 
we examine now the calculation of the electron-phonon coupling. With this purpose, 
we can expand the interaction between the electron and the ions $V_{ei}$ in the form

\begin{equation}
V_{ei}= \sum_{ij} V_{ei}({\mathbf r}_j - {\mathbf R}^0_i)  -\sum_{ij} {\mathbf Q}_i \nabla_j V_{ei}({\mathbf r}_j - {\mathbf R}^0_i ) + ....
\label{vex}
\end{equation}

\noindent where ${\mathbf Q}_i$ measures the separation of the ion
position  ${\mathbf R}_i$  from it equilibrium value ${\mathbf
R}^0_i$ i.e.  ${\mathbf Q}_i={\mathbf R}_i-{\mathbf R}^0_i $. The
first term in (\ref{vex}) represents the potential that  interacts
with the electron in the absence of lattice deformation. Then, the
information concerning the electron-phonon interaction is
contained in the second term \cite{phillips}.  We can separate the
interaction electron--ion in the form $V_{ei} = V_{s \sigma} +
V_{el} $, where $V_{s \sigma}$  corresponds to the Hund coupling
and $V_{el}$ includes all other possible electron--ion
interactions. Supposing that the position dependence of $V_{s
\sigma}$ is of the form $V_{s \sigma} = {\cal V}_{s
\sigma}({\mathbf r}_j - {\mathbf R}_i)( -J_H \sigma \cdot {\mathbf
S}_i )$, the second term in the 
expansion (\ref{vex}) may be put as

\begin{equation}
V_{ei2} = - \sum_{ij} {\mathbf Q}_i \nabla_j ( V_{el} ({\mathbf r}_j - {\mathbf R}^0_i ) -J_H \sigma \cdot {\mathbf S}_i   {\cal V}_{\sigma S}({\mathbf r}_j - {\mathbf R}^0_i ) ).
\label{vex2}
\end{equation}

\noindent  From the above expression, it becomes clear that Hund interaction can
 introduce large changes in the interaction electron phonon, principally  when
 $J_H$ is the strong coupling regime. To obtain a relationship involving $m$ and
 $g$, we shall assume that the form of $V_{el} ({\mathbf r}_j - {\mathbf R}^0_i )$ 
and ${\cal V}_{\sigma S}({\mathbf r}_j - {\mathbf R}^0_i )$ are similar. In the appendix,
we show that this approximation is adequate under certain conditions. 
Using this simplification we can write  
${\cal V}_{el} ({\mathbf r}_j - {\mathbf R}^0_i ) =  \xi V_{\sigma S}({\mathbf r}_j - {\mathbf R}^0_i ) $, 
where $\xi$ is a constant.  Inserting this equality in (\ref{vex2}), we obtain that

\begin{equation}
V_{ei2} = - (1- \xi J_H \langle \sigma_i \cdot {\mathbf S}_i \rangle ) \sum_{ij} {\mathbf Q}_i \nabla_j  V_{\sigma S}({\mathbf r}_j - {\mathbf R}^0_i ) ),
\label{vex3}
\end{equation}

\noindent where $J_h \langle \sigma \cdot {\mathbf S}_i \rangle$
is the spatial average of the Hund interaction. The interaction
electron-phonon takes the usual form, if we except the factor $(1-
\xi J_H \langle \sigma \cdot {\mathbf S}_i \rangle )$. The usual
electron-phonon interaction that does not depend on the
magnetization  is recovered when $J_H=0$.  In order to obtain the
approximate dependence of $g$ with $m$, we again follow the
method used in  \cite{kubo},\cite{anderson}  and consider a dimer
of two ions with  one itinerant electron. If we suppose that
initially the electron is at the ion $i=1$, in virtue of the
strong Hund coupling the spin is aligned with the spin of this
ion. When the electron jumps to the site $i=2$, it will interact
with the spin ${\mathbf S}_2$. In general $\sigma_1$ and ${\mathbf
S}_2$ will not be aligned. To obtain the value of  $J_h \langle
\sigma \cdot {\mathbf S}_i \rangle$ we average $J_h \langle
\sigma_1 \cdot {\mathbf S}_2 \rangle$ over all the states of the
total spin ${\mathbf S}_T = \sigma+{\mathbf S}_1+{\mathbf S}_2$
in the presence of the field $\lambda$. Due to the fact that
$\sigma$ is aligned with ${\mathbf S}_1$, these two spins can be
considered as an unique  spin with module $\bar{S} = S +
\frac{1}{2}$. Since $S = 3/2$ and $\sigma = 1/2$,   $ \langle
\sigma \cdot {\mathbf S}_2 \rangle$ =$  \langle { \bar{\mathbf
S}}_1 \cdot {\mathbf S}_2 \rangle /4 $. To calculate this average
we use the identity ${ \bar{\mathbf S}}_1 \cdot {\mathbf S}_2 =
\frac{1}{2} ( {\mathbf S}_T^2-{ \bar{\mathbf S}}_1^2-{\mathbf
S}_2^2)$. Then, the average is explicitly obtained from

\begin{eqnarray}
 \langle \sigma_1 \cdot {\mathbf S}_2 \rangle  =  \ \ \ \ \ \ \ \ \ \ \ \ \ \ \ \ \ \ \ \ \ \
\ \ \ \ \ \ \ \ \ \ \ \ \ \ \ \ \ \ \ \ \ \  \ \ \ \ \ \
\nonumber \\  \frac{\sum_{S_T = 1/2}^{\bar{S}_1+S_2 }  \sum_{M=-S_T}^{S_T} \langle SM | {\mathbf S}_T^2  -{ \bar{\mathbf S}}_1^2-{\mathbf S}_2^2 |SM \rangle   e^{M \lambda}}
{8 \sum_{S_T = 1/2}^{\bar{S}_1+S_2} \sum_{M=-S_T}^{S_T}  e^{M \lambda}}
\label{S}
\end{eqnarray}

\noindent where $\langle SM | {\mathbf S}_T^2-{ \bar{\mathbf S}}_1^2-{\mathbf S}_2^2) |SM \rangle = (S_T (S_T+1)-\bar{S}_2 (\bar{S}_2 +1)-S_1 (S_1 +1)) $, with  $\bar{S}_1 = 2$  
and $S_2 = 3/2$. Since $m$ and $\langle \sigma \cdot {\mathbf S}_i \rangle$ 
are functions of $\lambda$, we can obtain the relationship  between these 
quantities.  This dependence is shown in Fig. 1, and it can be  accurately
 approximated as $\langle \sigma \cdot {\mathbf S}_i \rangle =  \frac{1}{2} S \ m^{1.8}$.
  $\langle \sigma \cdot {\mathbf S}_i \rangle$ takes it maximum value  of 
$\frac{1}{2} S$ in the fully polarized state $m=1$, when $\sigma$ and
 ${\mathbf S}_i$ are aligned, and becomes zero in the disordered state $m=0$. 
Then, from  (\ref{vex3}), the dependence of the electron-phonon coupling with $m$ 
can be expressed as
\begin{equation}
g = g_0 (1- \alpha m^{1.8}) ,
\label{g}
\end{equation}

\noindent where $\alpha$  is a constant and $g_0$ is the value of
the e-ph coupling in  absence of magnetization. From this
expression, $\alpha$ can be expressed as $\alpha =
(g_0-g_{m=1})/g_0$, i.e. is the  variation of $g$ 
between the paramagnetic state ($m=0$) and the fully polarized
state ($m=1$) relative to $g_0$.   We notice that in previous works \cite{millis2},
the notion of an effective e-ph coupling that changes with $T$ was
already employed. However, the variations of $g$ considered here
are not only effective but real ones. As we shall show later, the
real variations of $g$ introduces important modifications in the
behavior of the system.

\section{Numerical results}

The numerical solutions are obtained minimizing free energy
$\cal{F}$ (\ref{F}) subject to the conditions (\ref{m}) and
(\ref{g}). In Fig. 2 we show the magnetization as a function of
$T$ for $\alpha = 0.4 $,  a doping level $ x =  1 - \langle n_i
\rangle = 0.25$ and different values of $g_0$ . Here a square DOS
bare band of bandwidth $D = 12 t$ was assumed. In general, the
effect of the e-ph coupling is to reduce the magnetic critical
temperature, for either of the cases in which $g$ is dependent or
independent of $m$. This occurs due to the narrowing band  effect
of the electron-phonon coupling, which reduces the efficiency of
the double exchange mechanism. However, when  $g$ is
$m$--dependent, the magnetic transition becomes discontinuous as
long as $\alpha$ is above a critical value $\alpha_{c}$ that depends
on the value of $g_{0}$ ( see Fig. 3 ). We notice that as $\alpha$
increases, both the critical temperature and the value of $m$ may
increase. This is because larger values of $\alpha$ reduces the
value of $g$ in the magnetic state, increasing the bandwidth and
favoring the DE mechanism. On the other hand, when the value of
$\alpha$ is increased, the critical value of $g_0$ for the ocurrence
of discontinuous transitions decreases. The minimal value of $g_0$ for 
the occurrence of
first order transitions  goes to infinity as $\alpha \rightarrow
0$, indicating that there are not discontinuous transition  when
$\alpha = 0$ ( $g$ independent of $m$) within our mean-field
treatment. In Fig. 4 we show the phase diagram in the ( $g_0$ ,
$\alpha$ ) space. The labelled regions correspond to: I)
continuous ferromagnetic-transitions, II) discontinuous
magnetic-magnetic transitions and III) discontinuous
ferromagnetic-paramagnetic transition. The three cases are
depicted in Fig. 3. Figs. 2, 3 show that the
magnetic transitions obtained with an $m$--dependent $g$ compares
well with the one obtained in experiments when $\alpha > 0 $ and $g_0$ is strong. 
At this point we remark
that the detailed form of the dependence of $g$ on $m$ does not
affect significatively the system behavior. If we employ a simple
linear dependence of the form $g = g_0 ( 1 - \alpha m )$, the phase  diagram is 
nearly the same as those showed in Fig. 4, although
there are some changes in the critical temperatures. Although we have not 
calculated the value of $\alpha$ from first principles, it
is worth mention that the first order transitions appear for any
value of $\alpha > 0$, if $g_0$ is  sufficiently  large.
Experimentally, it was found that the electron-phonon interaction
may undergo variations of order 150 \%  \cite{zhong}. If we
assume this level of variations, the value of $\alpha$ may be
taken between $0$ and $0.6$.

An important property of the manganites  is the strong dependence 
of their resistivity with the application of a magnetic field near $T_c$. 
In order to study if
this effect appears in the present model, we examine the 
dependence of the bandwidth with an applied  field. The average physical 
magnetization in $z$ 
direction is given by $M = g_s \mu_B N \langle S_z \rangle $, where $N$ is the 
number of ions. Thus, the energy due to the magnetic interactions is given by 
$H_M = N h m$, with $h=g_s \mu_B S H$. We then added 
the  term  $H_M$ to the free energy (\ref{F}) and obtained the mean-field 
solutions. The variations of $W$ are shown in figure 5. As it can be seen, when the 
transitions are discontinuous ($\alpha = 0.4$), there is a strong dependence of 
the bandwidth with $H$ near $T_c$. Since $W$ is proportional to conductivity, this 
means that there is a large increment of the conductivity with application of $H$. 
This effect is relatively weak when the transition is smooth ($\alpha = 0.1$).   
 
In many works it has been reported that the first order transitions in manganites are 
accompanied by phase separation. In order to analize this posibility  we ploted the 
free energy of the F and P solutions as a function of doping for constant 
temperature. Tendency to phase separation is established when the stability condition 
$(\frac{\partial ^2 \cal{F}}{\partial N^2})_{_T} \geq 0$ is not satisfied. Each solution 
satisfies this condition separately, but at the point at which 
the two free energies cross, 
the first transition takes place and the condition is not satisfied. We then search 
for bi-phase  solutions of the form: 
${\cal F} = N_1 {\cal F}_1(N_{e1} / N_1) + N_2 {\cal F}_2(N_{e2} / N_2)$, where $N_i$
 are the number of sites ocuppied by phase $i$ and  $N_{ei}$ the numbers of electrons in 
the volume occupied by phase $i$.  Thus, $N_1+N_2=N$ and $N_{e1}+N_{e2}=N_e$, 
being $N_e$ the 
total number of conduction electrons. ${\cal F}_1$ and ${\cal F}_2$ 
are the free energies for each phase, showed in Fig.6. Defining $a_1=N_1/N$, $z_1=N_{e1}/N_e$ and 
$n=N_e/N$, we can express the free energy per site as:
\begin{equation}
{\cal F}(T,n)/N = z_1 {\cal F}_1(T,n z_1 / a_1) + 
(1-z_1) {\cal F}_2(T,\frac {n (1-z_1)}{1-a_1}).
\end{equation}
We then obtained the volume fraction occupied by each phase 
minimizing ${\cal F}$ with respect to $z_1$ 
and $a_1$, with the restriction that the densitiy of electrons $n_i = N_{e1}/N_1$ must 
range between 0 and 1 in each phase. The volume fraction occupied by each phase is 
shown in figure 7, from which we can see that phase separation 
takes place around the critical conditions for the first order transitions.

As mentioned previously, the origin of the resistivity maximum
that appears near the critical temperature is still not well
understood. As already mentioned, this peak is obtained in the
DE-H model \cite{millis2},\cite{verges},\cite{dagotto}, when the
electron-coupling $\gamma$ takes values in a  relatively narrow
range. The calculations of the resistivity in these previous works
show that $\rho$  decreases  monotonically with $T$ if $\gamma <
1.2$. On the other hand, if $\gamma > 1.6$ , the resistivity
rapidly increases with  $1/T$. This last behavior is probably due
to the formation of polarons, whose resistivity exhibit such a
dependence with $T$. Although the spin scattering could contribute to develop a
maximum at $T_c$, the resistivity of the polarons increase so
rapidly that the peak is not observed.  In order to estimate the
variation of the resistivity due to polaron formation, we shall
use the expression for $\rho_{pol}$ in the strong coupling
adiabatic limit  ($\omega / t \rightarrow  0$)

\begin{equation}
\rho_{pol}(T)=  A T \exp(\frac{\epsilon_a}{k_B T})
\label{rpol}
\end{equation}
\noindent where $\epsilon_a \approx  \epsilon_p / 2  $ is the
activation energy \cite{salamon} and $\epsilon_p = g^2 \omega$ is the polaron binding
energy. From the Kubo's formula for resistivity it follows \cite{schubert} that 
$\rho \propto 1/t^2$. Then, we can write $A \propto 1/W^2$, where  $A$ is
 the parameter appearing in (\ref{rpol}). In various works it was found that
 this expression agrees
very well with the resistivity in manganites in the region $T >
T_c$ \cite{lu}, \cite{hartinger},\cite{banerjee}, revealing the
presence of small polarons in the paramagnetic phase. It is worth mention that 
in principle it is difficult to differenciate the polaronic resistivity from 
the  resitivity of an ordinary semiconductor, which has a similar dependence  to 
(\ref{rpol}) (without the prefactor $T$). However, in Ref. \cite{palstra} it has 
been shown that a band insulator theory cannot explain the differences between 
the gap value  obtained from thermopower and the one obtained from resistivy 
measurements. On the other hand, these differences may be explained assuming 
polaronic interactions.  

 We studied
the variation of $\rho_{pol}(T)$  using  (\ref{rpol}) with the
values of $g$ and $W$ obtained in the mean-field solutions. In order to consider 
a realistic value of $\omega$, we included the third term in (\ref{Hpro}) and minimized
 the free energy given by

\begin{equation}
{\cal F}=\Omega_e(\lambda) - N T [ \ln z(\lambda) - \lambda m(\lambda)] 
- \omega g^2(\lambda)  \sum_{i} n_i .
\label{F}
\end{equation}
In order to take realistic values of the parameters, we used the estimations of the
activation energy and phonon frequency  given in  \cite{banerjee}, 
$\epsilon_a = 0.08 $ eV, $\omega = 0.036 $ eV, which gives $g_0 = 2$. On the other hand, 
the bare bandwidth has been estimated in  \cite{millis} to be of order $2.5 $ eV. Since 
in our model $D=12 t$, this gives $\omega \approx 0.2 t$.  
In  Fig. 8  the values of log($\rho_{pol}$) for $g_0=2$
 $\alpha=0.4$ and $\omega = 0.2$ ($t=1$)are shown.  From this figure we can see
that a peak accompanied by a pronounced depletion of $\rho$ is
observed at $T = T_c$. This is caused by a rapid reduction of $g$
( in this case from $g = 2$ in the P state to $g=1.2$ in the F
state). As it can be seen from Fig. 8,  relatively small changes in
$g$ produce drastic changes in $\rho_{pol}$. Moderate variations
of $\gamma$ or other system parameters will not suppress the
presence of these peaks. These  appear in region III of
Fig. 4. Of course, we have not considered the scattering of the
electrons by the spins. We expect that this effect will superpose
to the polaronic ones and contribute to increase the resistivity
peak.

We note that the reduction of $g$ allows a transition from small
polarons (large coupling) in the P state to large polarons (small
coupling) in the F state. This transition has been reported by a
number of groups \cite{billinge}, \cite{kim}, \cite{lanzara}. The
variation of the resistivity for $T < T_c $ also supports
reduction of the electron-phonon coupling  in the ferromagnetic
region. It has been shown  that resistivity in the
ferromagnetic phase of the manganites can be well described with a
dependence of the form $\rho = \rho_0 + A T^2 + B T^{4.5}$, where
the first term is a constant, and the two last terms correspond to
electron-electron interaction and electron-magnon scattering
respectively  \cite{snyder},\cite{lu}.  A kind of transition
between two polaronic regimes, as the one found here, could
explain the small influence of polaronic effects in the   
resistivity below  $ T_c$. We comment that we do not attempt to describe 
accurately the resistivity of the F phase with (\ref{rpol}), because it is 
valid for the strong coupling regime. Thus, the resitivity of Fig. 8 describes the
transition between two phases with different e-ph coupling, both in the
strong coupling case. 
It is worth mention that 
the form of this curve is similar to the one obtained in $La_{0.85}Sr_{0.15}MnO_3$. At 
$T_c = 240 K $, this compound experiments a transition from a P insulator to an F 
insulator  \cite{tokura} (see inset in Fig. 40 of Ref.  \cite{salamon}).

A recent  study of the quasiparticle
excitation spectrum of  $La_{0.77}Ca_{0.23}MnO_{3}$ using scanning
tunnelling
microscopy by S. Seiro  {\it et al}  revealed that the spectra presents
a polaronic gap in the F and P phases. However, the gap is reduced when
the system enters in the F phase. Since the polaronic gap measured
in the experiment is of magnitude $\epsilon_p$, the dependence of the gap
with the temperature is compatible with the reduction of the e-ph interaction in
the F state that is obtained in the present model.


\section{Conclusions}

In this work,  we
study a DE-H model with an electron-phonon interaction that depends
on the magnetic ordering. By a simple argument, examining the
calculation of the electron-phonon coupling, we show that $g$ is
affected by the magnetic interactions. This occurs principally
when the Hund coupling is strong, which is precisely  the regime
that is considered to be relevant for the  manganites. Introducing
a field parameter to control the magnetization and making some
assumptions about the spatial form of the electron-ion
interaction, we calculated a dependence of $g$ with the
magnetization $m$. When $\alpha$ ($ \sim  dg/dm $) is above a
critical value that depends on $g_0$, the  F-P transitions
becomes discontinuous. 
This is not only because the DE mechanism and the polaron
formation  are  affected mutually in an indirect form, but also
because there is a interplay between these two mechanisms. In
the spin disordered state, the electron--phonon interaction is
enhanced, favoring the appearance of small polarons. In turn, the
polaronic effect reduces the electrons mobility decreasing the
efficiency of the DE mechanism. This highly non linear effect
induces a sharp magnetic transition. 
These first order transitions may occur
directly between  the F to P states or between two F
states.  The last case  appears only in a narrow region of the
parameter space and probably will not be present in a more refined treatment of
the model  than the mean-field approach considered
here. On the other hand, we can expect that a crossover from
smooth to sharp transitions will persists in region II if
fluctuations are included.  The abrupt transitions of region III
(Fig. 4) have a shape that  resembles the
magnetic transitions experimentally observed in the manganites.
 We estimated the changes in the polaronic resistivity using the expression 
of $\rho_{pol}$ in the adiabatic limit. A strong peak in the resistivity of
 polaronic origin appears at the critical temperature when the transitions are 
abrupt. This is very robust and appears in region III of the phase diagram of Fig. 4. 
 This  allows to explain the presence of the resistivity maximum in a series of 
compounds, for which $\gamma$ and other properties may be different. For some
 values of the parameters, the discontinuous transitions are accompanied by a 
crossover from large polarons in the FM phase to small polarons in the PI phase, 
in agreement with several experimental observations \cite{billinge,kim,lanzara}.
In summary, the obtained results show that the DE-H model with an electron-phonon 
coupling that varies with the magnetization reproduces some properties of the 
manganites that are still not well understood.   
It is not our intention  to neglect the importance of other factors such as structural
 disorder or phase separation in order to explain the manganites behavior, but
instead of proposing 
another mechanism that could be important to complete the description of 
these systems.

\begin{center}
{\sc Acknowledgments}
\end{center}

We thank the {\it Programa de Desarrollo de las Ciencias B\'asicas} (PEDECIBA), Uruguay, for partial financial support.

\begin{center}
{\sc Appendix}
\end{center}

The spin dependent part of the electron-ion interaction is an exchange interaction, 
given by
\begin{equation}
J(R)= \int \! \! \! \!  \int \phi_1(r) \phi_2(r') V(r'-r) \phi_2(r') \phi_1(r) dr dr'.
\end{equation}
The orbital functions $\phi_i(r)$ are centered on different points separated by
 a distance $R$. In order to study the dependence of this  interaction  with $R$, 
we follow Campbell {\em et al}  \cite{campbell} and assume that the e-e interaction is 
a screened potential of the form $V(r'-r)=A \exp(-a|r-r'|)$, being $r-r'$ the distance 
between the electrons. We also assume  that the functions $\phi_i$ are of the 
form $\phi(r)=C \exp(- \kappa r)$. Here $1/a$ is the screening length and 
$1/\kappa$ is the localization length. We evaluated $J$ for different values of
 $a$, $\kappa$ and $R$. We obtained that when $a$ and $\kappa$ are nearly equal, 
the variation of $J(R)$ follows the variation of $V(R)$. When $a > \kappa$, $J(R)$ fall 
more rapidly than $V(R)$. In the case $a < \kappa$, the opposite occurs. 
Then, the assumption that $J(R)$ has the same spatial dependence than $V(R)$, it is valid 
for some particular physical conditions.    
On the other hand, it is worth mention that the overall results do not depend strongly 
on the functional dependence of $g$ with $m$. Thus, this assumption   was performed  
to allow an analytical calculation, but does not introduce results which disappear 
if other dependence is adopted.

\ \ \ \

FIGURE CAPTIONS

 \ \ \ \ \ \

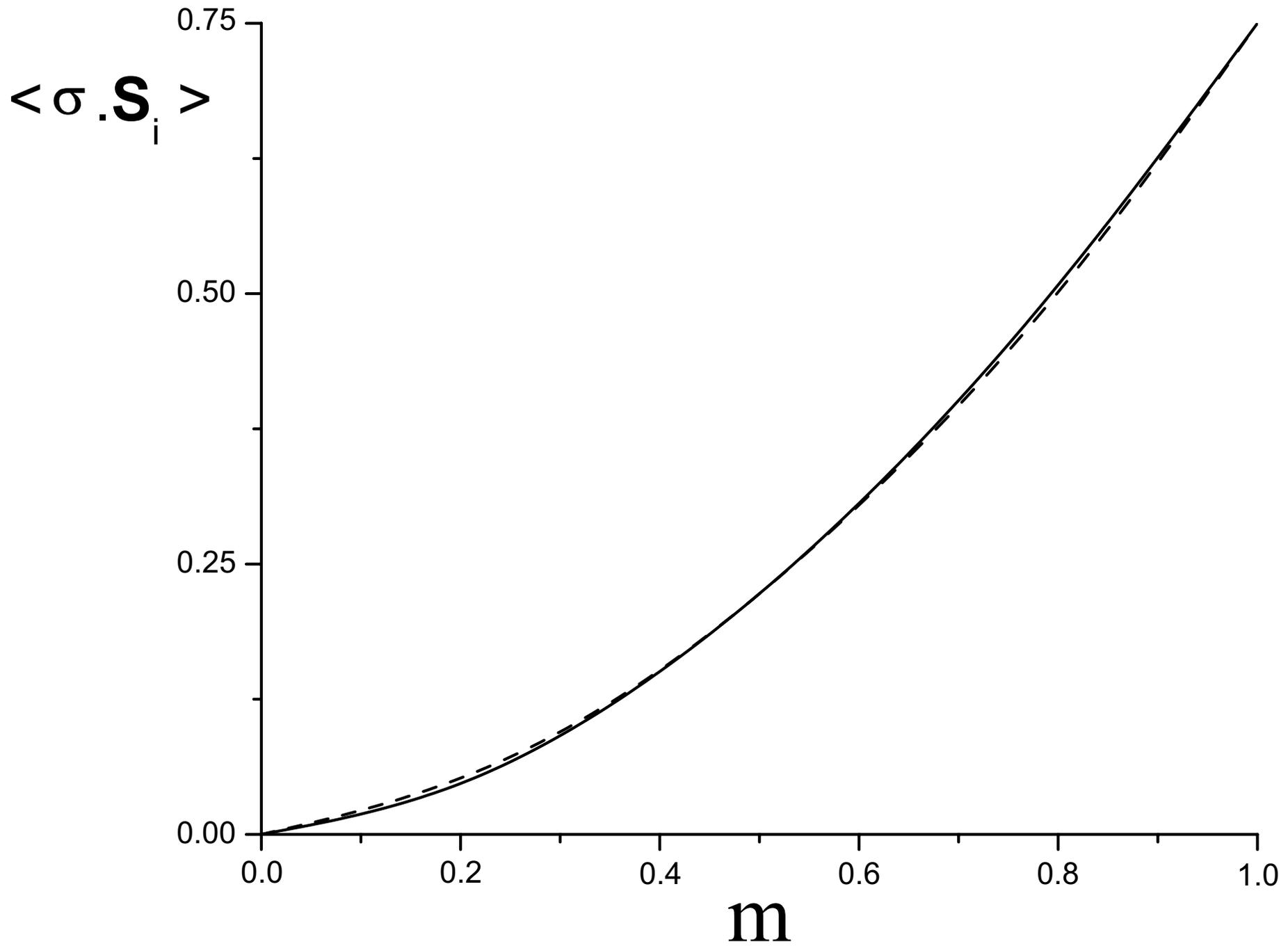
\noindent Figure 1:  Dependence of $\langle \sigma \cdot {\mathbf
S}_i \rangle$ with the magnetization. The continuous line
represents the values obtained with (\ref{m}) and (\ref{S}). The
dashed line corresponds to the approximate expression $\langle
\sigma \cdot {\mathbf S}_i \rangle = 3/4 m^{1.8}$.

 \ \ \ \ \ \

\noindent
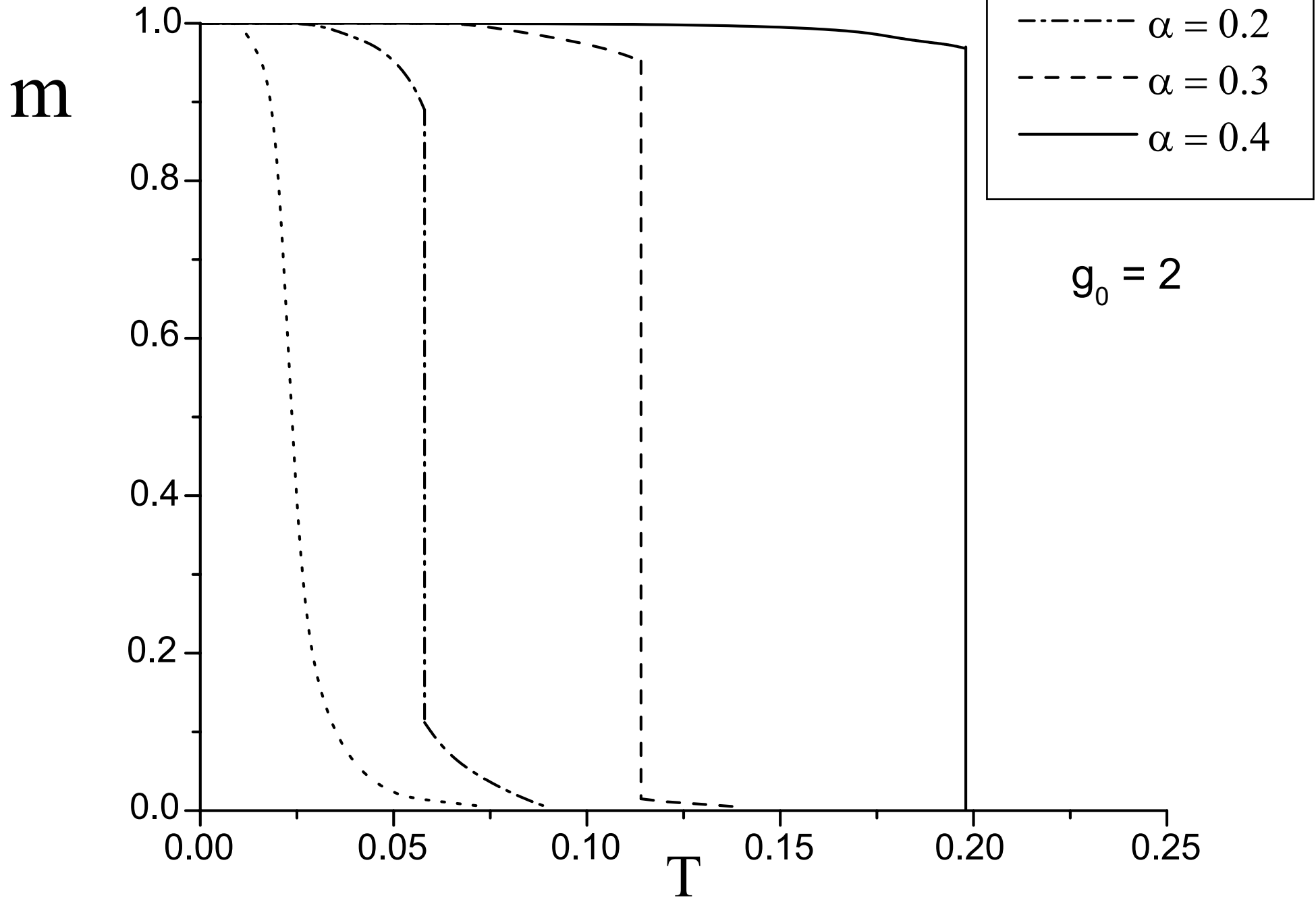
Figure 2: Magnetization as a  function of the temperature for $g_0 = 2 $ and 
$\alpha = 0.1, 0.2, 0.3, 0.4$. Temperature is measured in units of $t$.

\ \ \ \ \ \

\noindent
Figure 3: Magnetization as a function of the temperature for $\alpha = 0.4$ and $g_0 = 1, 1.5, 2$.

\ \ \ \ \ \

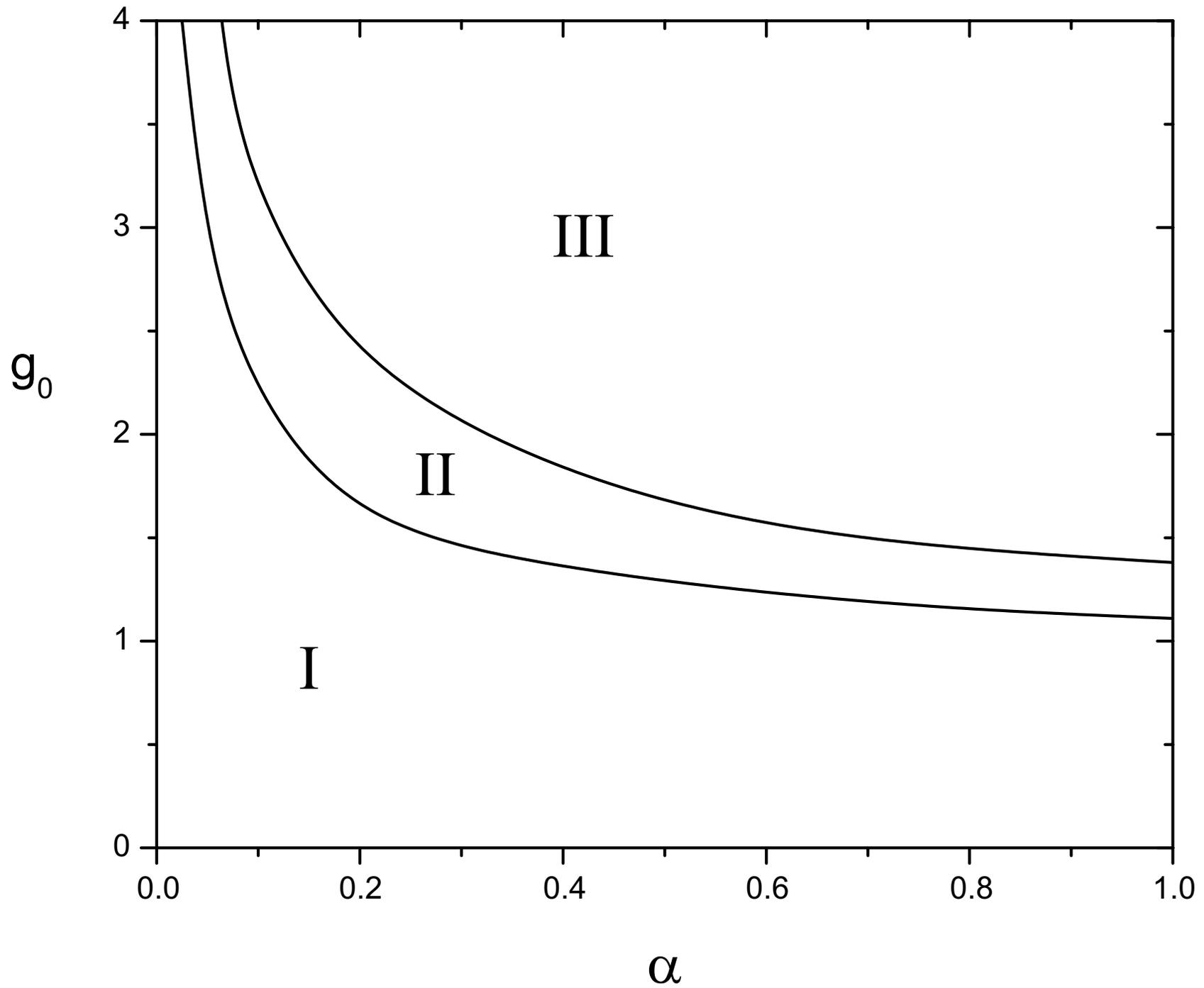
\noindent Figure 4: Phase diagram of the DE-H model with $g=
g_0(1-\alpha m^{1.8})$. The labelled regions represent: I)
continuous magnetic transitions, II) discontinuous F-F
transitions and III) discontinuous F-P transitions.

 \ \ \ \ \ \

\noindent
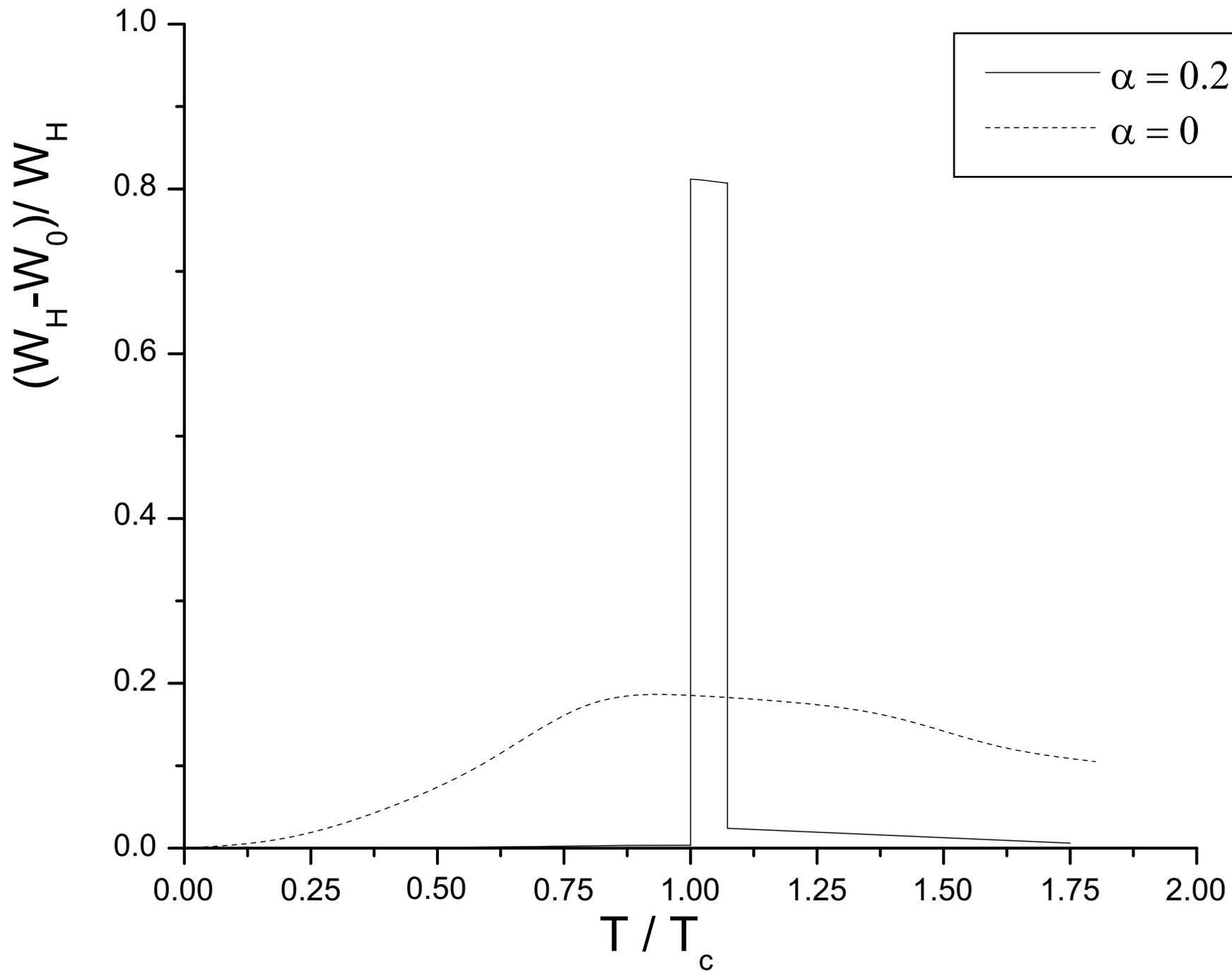
Figure 5: Percentual variation of the bandwidth with application of magnetic field, 
calculated as $(W(H)-W(0))/W(H)$, for $g=2$,  $\alpha = 0 , 0.2$ and $h=0.02$. 
$W(H)$ and $W(0) $  denote the bandwidths 
with and without magnetic field. 

 \ \ \ \ \ \

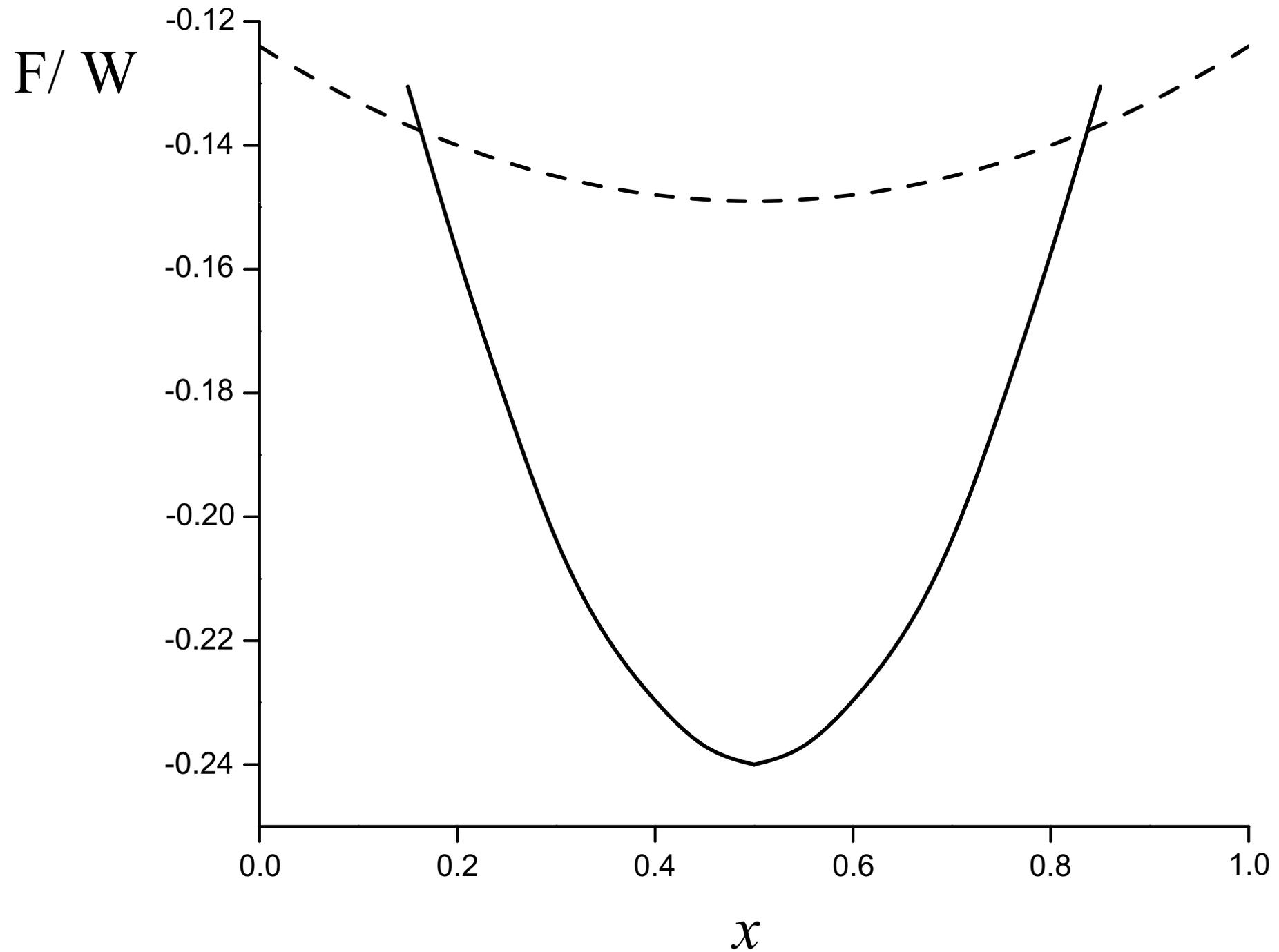
\noindent Figure 6: Free energies per site over the bandwidth, for $g=2$,
$\alpha=0.2$ and $k_B T$ =0.05 . The solid line corresponds to the F state
and the dashed line to the P one.
For these parameter values, phase separation takes place for doping value ranges 
 $0< x < 0.3$ and $0.7<x<1$.

\ \  \  \ \ 

\noindent
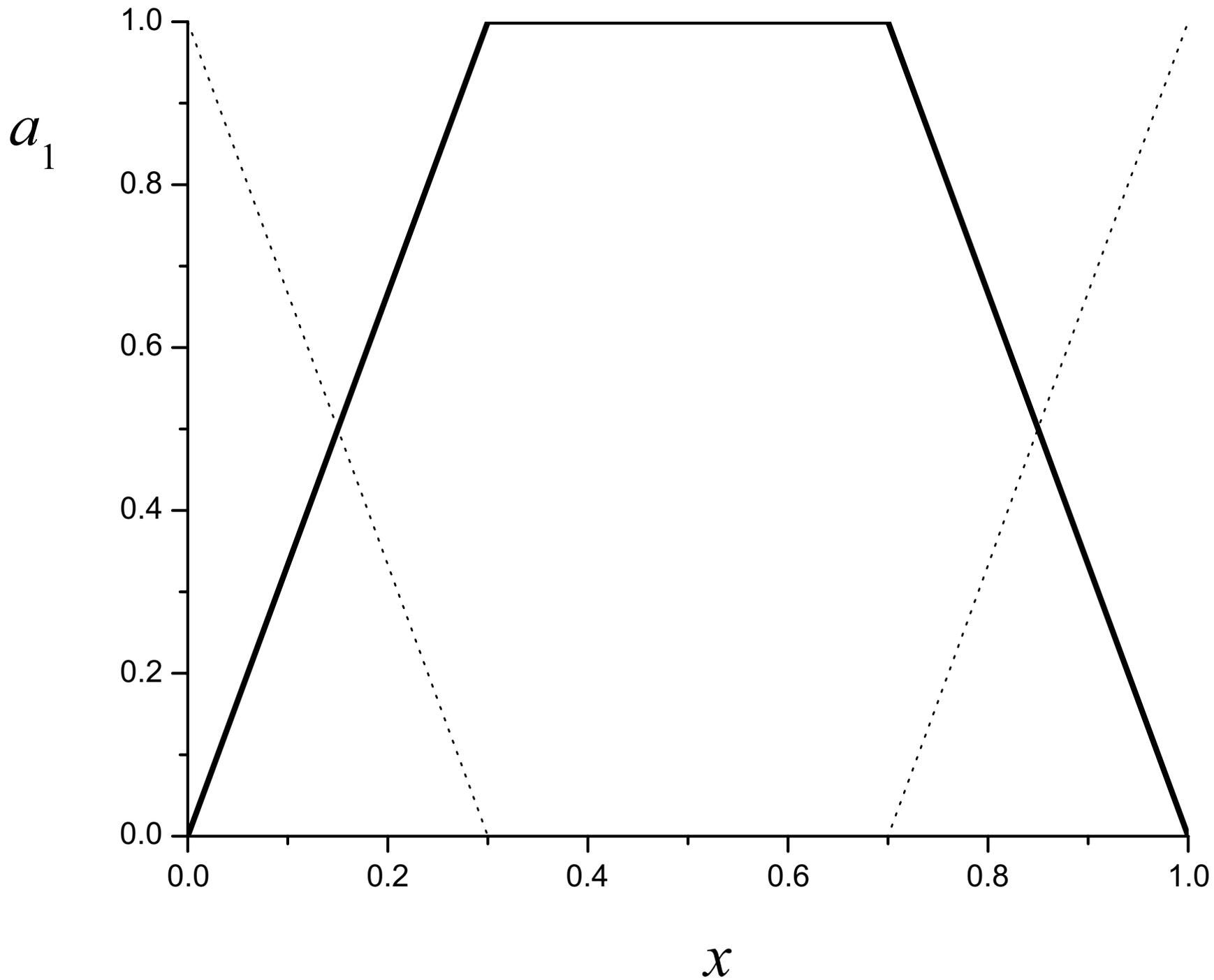
Figure 7: Volume fraction occupied by  F (solid line) and P
(dashed line) regions, for  $g=2$, $\alpha=0.2$ and $k_B T$ =0.05 .

\  \  \  \  \ 

\noindent
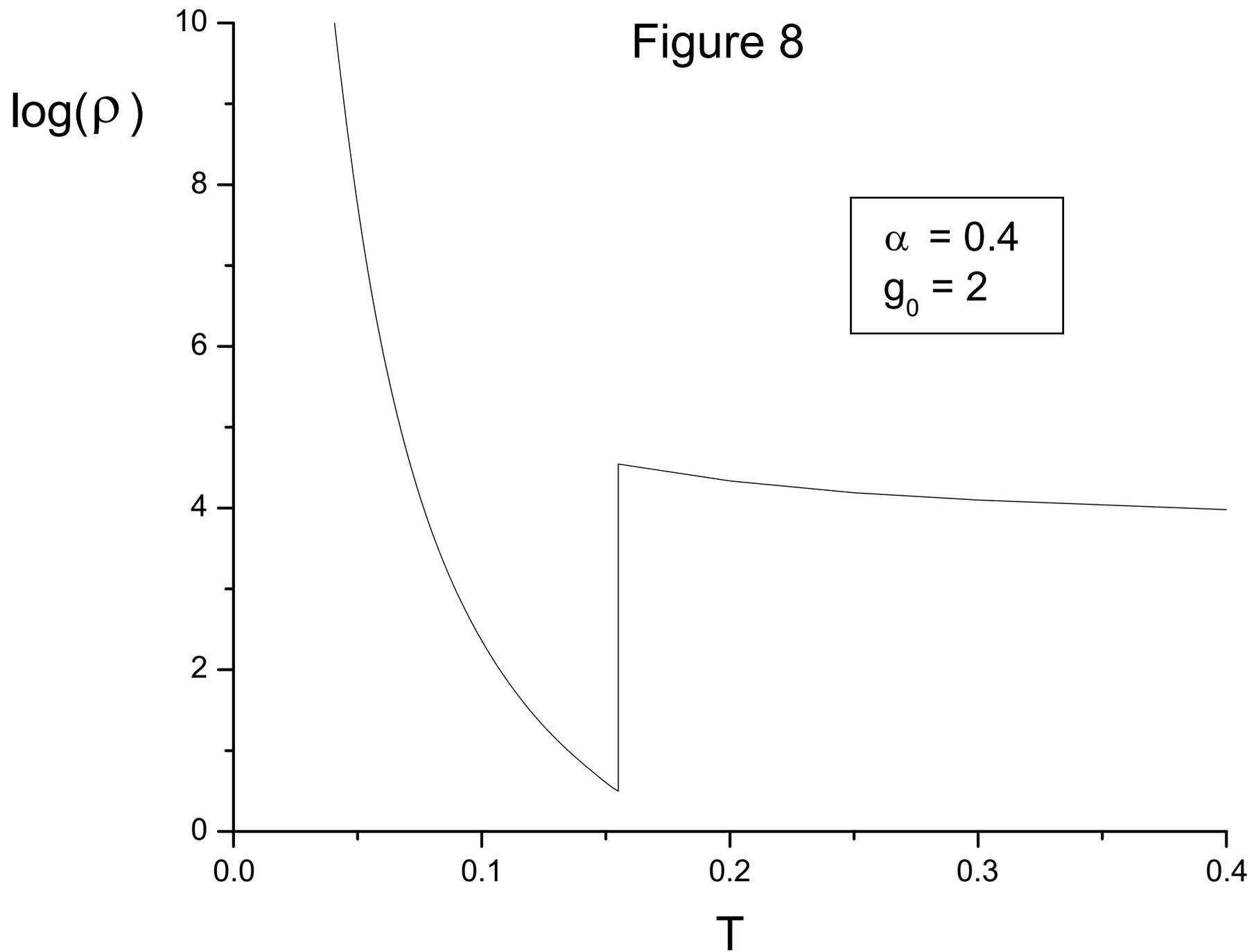
Figure 8: Logarithm of the polaronic part of the resistivity obtained from the 
expression $\rho_{pol}(T)=  \frac{B T}{W^2} \exp( \frac{g^2 \omega}{k_B T})$, 
using the values of $g$ and $W$ obtained in the mean-field solutions for $g_0 = 2$,
 $\omega = 0.2$  and $\alpha = 0.4 $ ($B$ is a constant).

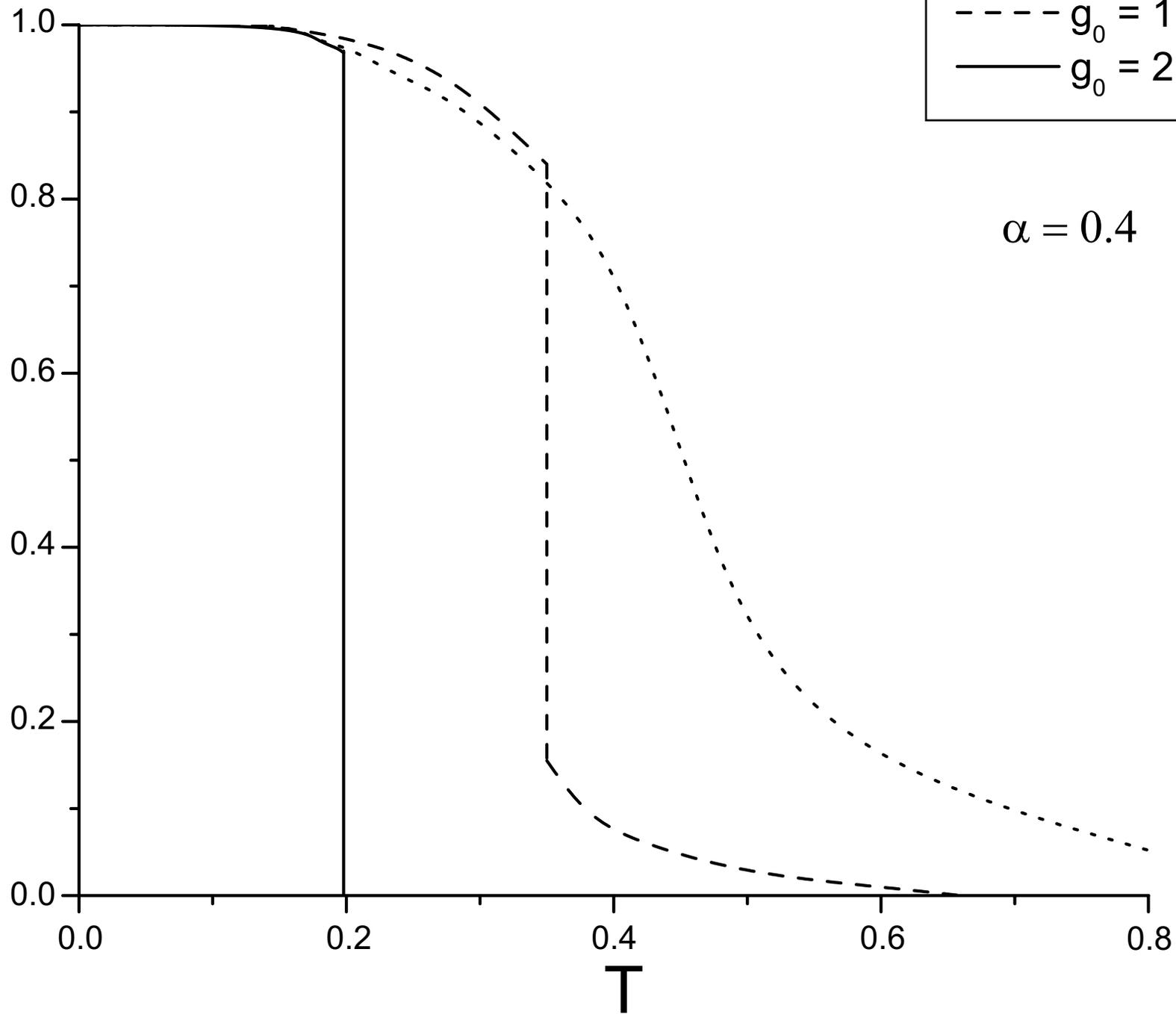
\end{document}